\documentclass{emulateapj}
\usepackage{textcomp}
\usepackage{mathtext}
\usepackage{amssymb}

\newcommand{\fermi}{\textit{Fermi}}
\newcommand{\gr}{$\gamma$-ray}
\newcommand{\snr}{SN~1006}

\begin{document}

\title{The Likely \textit{Fermi} detection of the supernova remnant SN~1006}

\author{Yi Xing\altaffilmark{1}, Zhongxiang Wang\altaffilmark{1},
Xiao Zhang\altaffilmark{2}, \& Yang Chen\altaffilmark{2,3}}

\altaffiltext{1}{\footnotesize 
Key Laboratory for Research in Galaxies and Cosmology,
Shanghai Astronomical Observatory, Chinese Academy of Sciences,
80 Nandan Road, Shanghai 200030, China}

\altaffiltext{2}{\footnotesize Department Astronomy, Nanjing University,
163 Xianlin Avenue, Nanjing 210023, China}

\altaffiltext{3}{\footnotesize 
Key Laboratory of Modern Astronomy and Astrophysics,
Nanjing University, Ministry of Education, Nanjing 210023, China}

\begin{abstract}

We report the likely detection of \gr\ emission from the northeast shell
region of the historical supernova remnant (SNR) SN~1006. Having analyzed 
7 years of \textit{Fermi} LAT Pass 8 data for the region of SN~1006, we
found a GeV gamma-ray source detected with $\sim 4\sigma$ significance. 
Both the position and spectrum of the source
match those of HESS J1504$-$418 respectively, which is TeV emission 
from SN~1006. Considering the source
as the GeV \gr\ counterpart to SN~1006, the broadband spectral energy 
distribution is found to be approximately consistent with the leptonic 
scenario that has been proposed 
for the TeV emission from the SNR. Our result has likely confirmed 
the previous study of the SNRs with TeV shell-like morphology: SN~1006 is one 
of them sharing very similar peak luminosity and spectral shape.
\end{abstract}


\section{Introduction}

As one of the few supernova remnants (SNRs) that were
historically recorded by people from different countries or continents
\citep{sg02,ste10}, 
\snr\ is of great interest and has been
extensively studied at multiple energy bands.
It is located far away from the Galactic plane, with a Galactic latitude
of $\sim$14\fdg5, in a relatively low ambient-density
($\sim 0.085$\,cm$^{-3}$; e.g., \citealt{kat+09}) environment.
The long-term proper motion measurements of the shock front at optical narrow
band, combined with the expanding velocity, implies a distance of
$\sim$2.2 kpc for the SNR \citep{wgl03}.
Multiwavelength emission from \snr\ shows that the remnant has a 
diameter of 30\arcmin (or 19 pc at 2.2 kpc), with two main lobes located
at northeast (NE) and southwest (SW) parts of the SNR's disk-like region.
While the interior of the SNR is dominated by thermal emission (e.g.,
\citealt{uyk13}), the shell is dominated by synchrotron emission 
which is bright in radio \citep{rg86} and 
hard X-ray \citep{rot+04,win+14} bands.

SNRs are considered to be the main sites in the Milky Way for producing
cosmic rays with energies up to a few 10$^{15}$\,eV.
Charged particles are accelerated in their shock fronts due to 
the diffusive shock acceleration mechanism (e.g., \citealt{be87}). 
As for \snr, in addition
to hard X-rays that indicate high-energy electrons accelerated to
100\,TeV in the shock front \citep{koy+95}, very high energy 
(VHE; $>100$\,GeV) emission
was also detected with the High Energy Stereoscopic System (HESS). The
two HESS sources J1504$-$418 and J1502$-$421 correspond to
the NE and SW shell regions respectively \citep{ace+10}, with the former
approximately 50\% brighter than the latter.  
Given these, GeV \gr\ emission from \snr\ has been
searched in observations with the Large Area Telescope (LAT) onboard
Fermi Gamma-ray Space Telescope (Fermi).
Using 3.5 and 6 years of \fermi\ LAT data, only upper limits have been 
obtained by \citet{af12} and \citet{ace+15}, respectively. 
Combining the TeV spectrum with the GeV upper limits, a leptonic scenario,
in which \gr\ photons arise from the inverse Compton (IC) scattering process,
is favored for the \gr\ emission \citep{ace+10,af12,ace+15}. 

The \fermi\ upper limits already tightly constrain the models typically 
considered for young SNRs (e.g., Cas A: \citealt{ace+10}, 
\citealt{ac10}; Tycho: \citealt{gio+12})
or for SNRs having a TeV shell-like morphology \citep{ace+15}.
Evidence for the interaction with a HI cloud in the SW limb region of
\snr\ has been reported by \citet{mic+14}. 
With the release of the best \fermi\ LAT dataset (Pass 8 data) in early 2015
and the accumulation of 7 years data, detailed analysis of the \gr\ emission
from \snr\ is thus warranted.  In this paper we report our analysis of 
the \fermi\
LAT data of the \snr\ region and the likely detection of \gr\ emission in
0.15--300 GeV energy range from the NE region of the SNR.

\section{Data Analysis and Results} 
\label{sec:ana}

\subsection{\textit{Fermi} LAT Data}

LAT is a $\gamma$-ray imaging 
instrument that scans the whole sky every three hours and is basically 
conducting long-term \gr\ observations of GeV sources \citep{atw+09}.
For this analysis we selected \textit{Fermi} LAT Pass 8 events in the energy range from 150 MeV to 300 GeV
centered 
at the SIMBAD position of \snr, which is 
($\alpha_{J2000}$, $\delta_{J2000}$) = (15$^{\rm h}$02$^{\rm m}$22$\fs$1, $-$42$^{\circ}$05$'$49$\farcs$0),
obtained by \citet{wo90}. 
The events below 150 MeV were excluded to reduce the effects of the Galactic 
background and the relatively large uncertainties of the instrument response 
function of the LAT in low energy range. 
The time period of
the LAT data is from 2008-08-04 15:43:36 (UTC) to 2015-09-24 00:03:16 (UTC). 
Following the recommendations of the LAT 
team\footnote{\footnotesize http://fermi.gsfc.nasa.gov/ssc/data/analysis/scitools/}, 
we included those events with zenith angles less than 90 degrees, which 
prevents the Earth's limb contamination, and 
excluded the events with quality flags of `bad'. 

\subsection{Source Detection}
\label{subsec:sd}

We included all sources within 20 degrees centered at the position of 
\snr\ in the \textit{Fermi} LAT 4-year catalog \citep{3fgl15}
to make the source model. The spectral forms of these 
sources are provided in the catalog.  Spectral parameters of the sources 
within 5 degrees from \snr\ were set as free parameters, and 
the other parameters were fixed at their catalog values. 
The background Galactic and extragalactic diffuse 
emission were also added in the source model with the spectral model 
gll\_iem\_v06.fits and the file iso\_P8R2\_SOURCE\_V6\_v06.txt, respectively, 
used.  The normalizations of the diffuse components were set as
free parameters.

\begin{figure}
\centering
\includegraphics[scale=0.25]{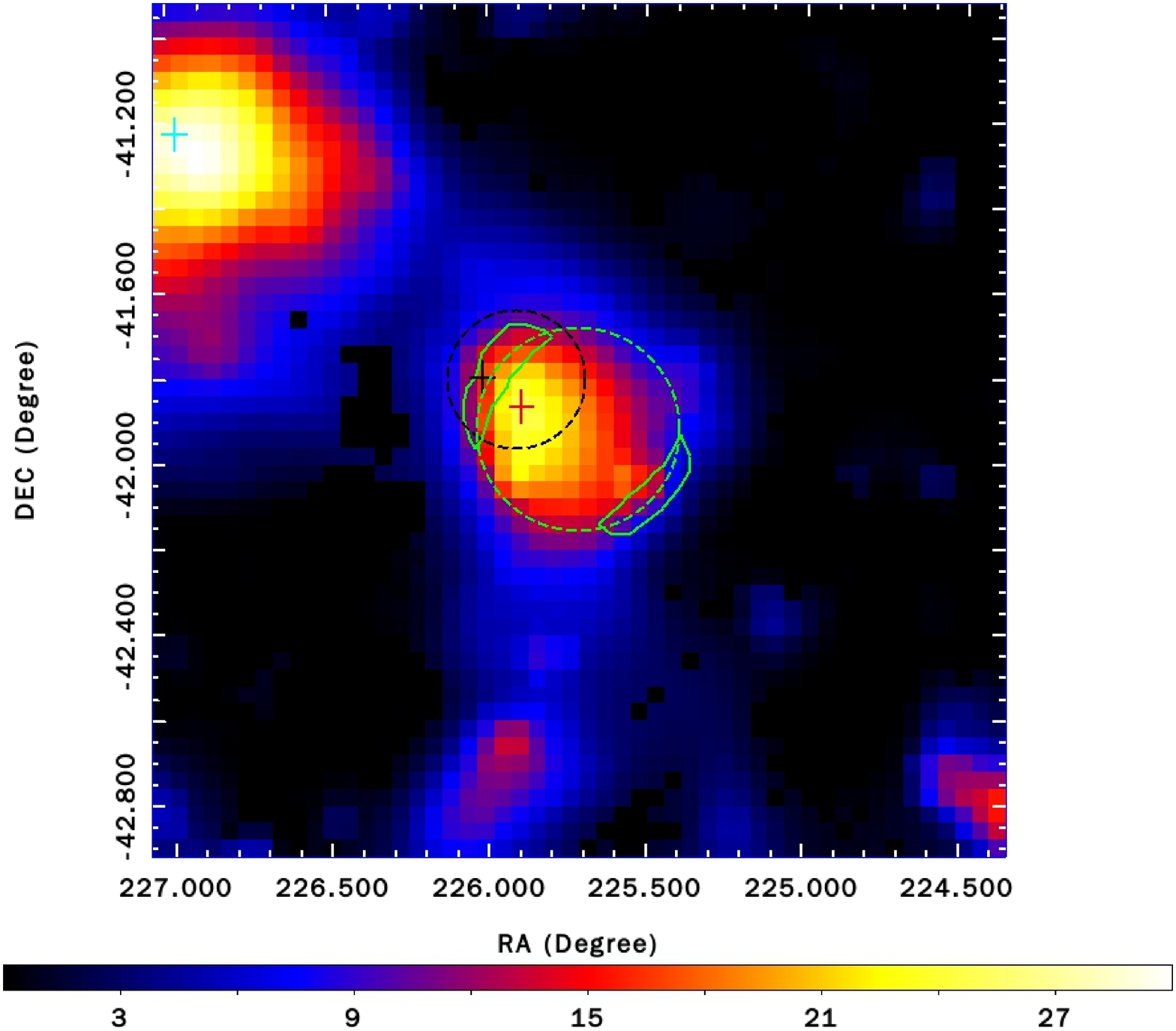}
\includegraphics[scale=0.25]{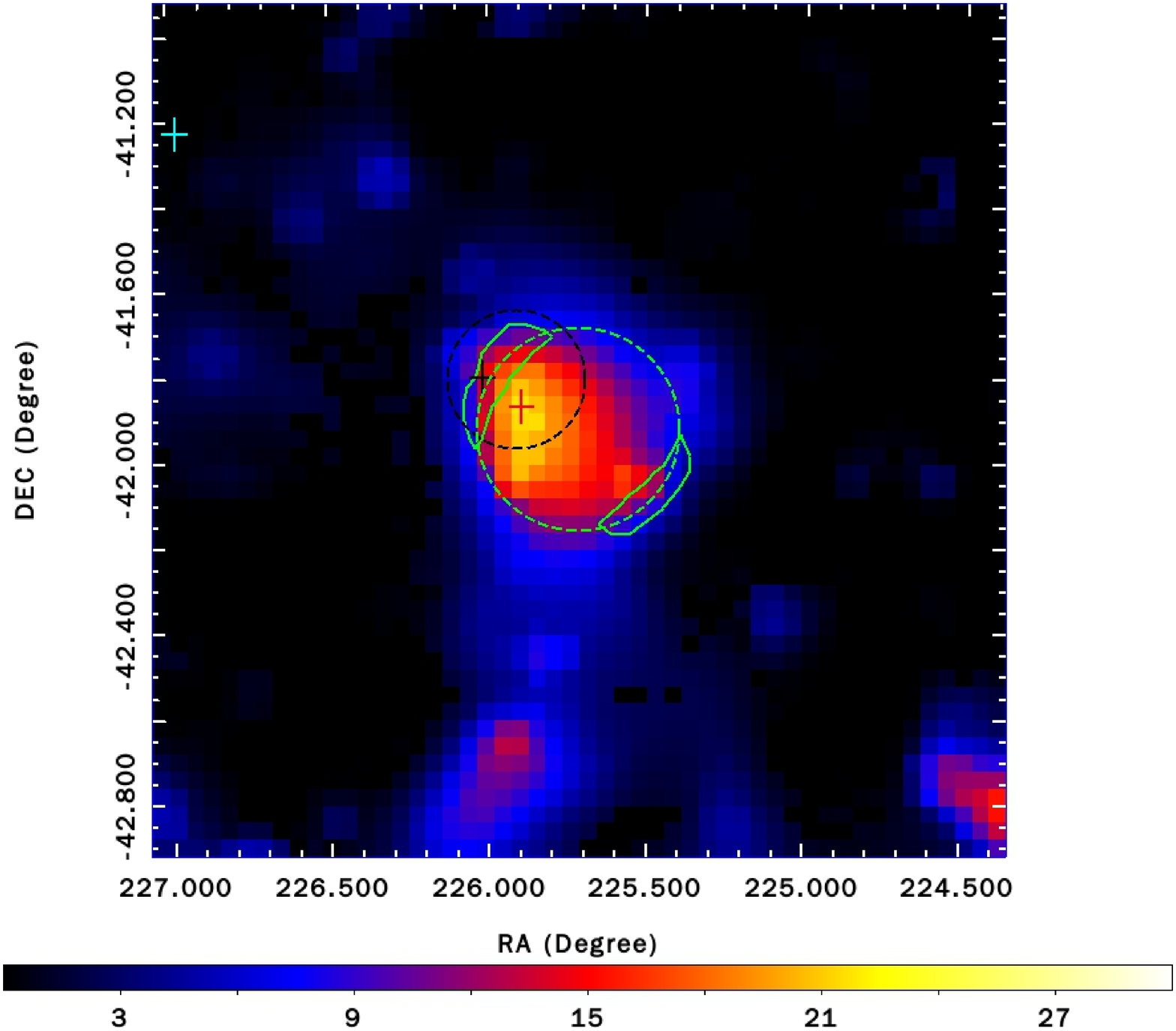}
\caption{TS maps of the $\mathrm{2^{o}\times2^{o}}$ region
centered at the X-ray center of \snr\ in the 1--300 GeV band, with
the green dashed circle indicating the X-ray region of \snr\ and the green
contours indicating the 2.0--7.2 keV X-ray intensity measurements of the NE and SW limbs
of \snr\ (at a level of 19 counts s$^{-1}$; obtained from the XMM-Newton map rebinned to the $\sim$0\fdg04 pixel$^{-1}$ scale).
The image scale of the maps is 0\fdg04 pixel$^{-1}$. All catalog sources were 
considered and removed. The non-catalog source detected in
the top left corner (the {\it top} panel, indicated by a cyan cross) was removed in the {\it bottom} panel.
The dark and red crosses mark the positions 
of HESS J1504$-$418 and QSO J1504$-$4152, respectively. The dark dashed circle 
marks the 2$\sigma$ error circle of the best-fit position obtained for 
the excess emission.
\label{fig:tsmap}}
\end{figure}

We first performed standard binned likelihood analysis to 
the LAT data 
in the $>$1 GeV band using the LAT science tool \textit{gtlike} 
in the science tools software package {\tt v10r0p5}. 
The Instrument Response Functions (IRFs) of P8R2\_SOURCE\_V6 were 
used.
Test Statistic (TS) maps obtained in this higher energy range would likely be better resolved 
in a possibly crowded region and source positions be better determined as well.
With the fitted source model, we calculated the binned 
TS map (using \textit{gttsmap} in the \fermi\ software package) 
of a $\mathrm{2^{o}\times2^{o}}$ region centered 
at \snr. All sources in the source model were considered and removed.
The obtained residual TS map of the source region is shown in the top panel of
Figure~\ref{fig:tsmap}. Excess emission in the \snr\ region was
detected with a maximum TS value of $\sim$25. 
The TS value at a given position is a measurement of the
fit improvement for including a source, and is approximately
the square of the detection significance of the source \citep{1fgl}.
Therefore, the excess was detected with $5\sigma$ significance.
We noted that no catalog sources are within the square region,
but there is an additional source in the top left corner of the TS 
map.
 
We investigated if the nearby additional source might contaminate 
the detection of the excess emission in the \snr\ region.
We ran \textit{gtfindsrc} in the LAT software package, 
and determined its position:
($\alpha_{J2000}$, $\delta_{J2000}$) = (227\fdg0, $-$41\fdg2) 
with 1$\sigma$ nominal uncertainty of 0\fdg2.
Adding the nearby source in the source model 
as a point source with power-law emission at its best-fit position,
we re-performed the binned likelihood analysis. 
The TS map with this nearby source considered in the source model is shown in the bottom panel of
Figure~\ref{fig:tsmap}.
Now with the nearby source totally removed, the maximum TS value for the 
excess emission in the \snr\ region was 
$\sim$22, still significant. In the following analysis the nearby source was considered in the source model. For the confirmed excess source, we determined its position, which
is ($\alpha_{J2000}$, $\delta_{J2000}$) = (225\fdg9, $-$41\fdg8) 
with 1$\sigma$ nominal uncertainty of 0\fdg1. The VHE
source HESS J1504$-$418 (\citealt{ace+10}, marked in Figure~\ref{fig:tsmap}) is 0\fdg08 from the best-fit position and within the 1$\sigma$ error circle.

We then investigated whether the confirmed excess source is point-like or extended.
Using both point-source and uniform disk models with
power-law spectra at the best-fit position, we performed likelihood analysis
to the data in 1--300 GeV energy range.
The radius for the uniform disk was set in a range of
0\fdg1--0\fdg5 with a step of 0\fdg1.
The spectral parameters of the sources within 5
degrees from \snr\ were set as free parameters, and all
other parameters in the source model were fixed at their catalog
values. No significant extended emission
was detected. The TS$_{ext}$ values, calculated from
TS$_{disk}$ $-$ TS$_{point}$ (see, e.g., \citealt{lan+12}),
were smaller than 0. We included the excess source as a point source in the source model with power-law emission and performed likelihood analysis in 0.15--300 GeV band. The photon index $\Gamma= 1.9\pm$0.3 and the photon flux $F_{0.15-300}= 7\pm 2\times 10^{-10}$ photons~s$^{-1}$\,cm$^{-2}$ were obtained, with a TS value of 15.

\subsection{Variability Analysis}
\label{subsec:lv}

In addition to the VHE source HESS J1504$-$418, there is an another known source located in the 2$\sigma$ error circle of the best-fit position, that is the quasar QSO J1504$-$4152 \citep{wl97}. 
Its position is marked in Figure~\ref{fig:tsmap}.
Given that active galactic nuclei (AGN) are the dominant source class 
detected by \fermi\ LAT \citep{3fgl15}, we thus 
searched for long-term variability for the purpose of checking any possible
association between QSO J1504$-$4152 and the excess source.
We calculated the variability 
index TS$_{var}$ for the point source in \snr\ region with 87 time 
bins (each bin constructed from 30-day data) in the energy range 
of 0.15--300 GeV, following the procedure introduced in \citet{nol+12}. 
If the flux is constant, 
TS$_{var}$ would be distributed as $\chi^{2}$ with 86 degrees of freedom. 
Variable sources would thus be identified with TS$_{var}$ larger than 119.4 
(at a 99\% confidence level in the $\chi^{2}$ distribution; see
\citealt{nol+12}).
The computed TS$_{var}$ for the source is 48.9, 
corresponding to a $<$1\% confidence level for a variable
source. The value indicates that there was no signifiant 
long-term variability observed in the \gr\ source.

\subsection{Spectral Analysis}
\label{subsec:sa}

\begin{figure}
\centering
\includegraphics[scale=0.3]{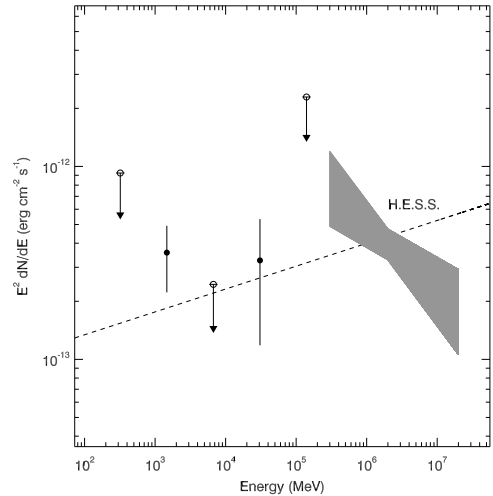}
\caption{\fermi-LAT \gr\ spectrum of the excess source at the best-fit position 
in the \snr\ region.
The 0.15--300 GeV power-law fit to the source is shown as
the dashed line. The grey area marks 
the power-low 
spectrum for the NE region of \snr, obtained with HESS.
\label{fig:spectra}}
\end{figure}

Considering the excess emission at the \snr\ region as a
point source at the best-fit position, we extracted its $\gamma$-ray spectrum 
by performing likelihood analysis to the LAT data 
in 5 evenly divided energy bands in logarithm from 0.15--300 GeV.
The excess emission was modelled with a power law in each energy band. 
The fluxes obtained in this way are not dependent on the emission model
assumed for a source, providing a good
description for the \gr\ emission of the source. 
In the extraction, the spectral normalizations of the 
sources within 5 degrees from the central position of \snr\ were set as 
free parameters, 
while all the other parameters of the sources were fixed at the values 
obtained from the above maximum likelihood analysis.
The obtained spectrum is plotted in 
Figure~\ref{fig:spectra}, where we kept the data points with TS greater 
than 9 (corresponding to the detection significance of 3$\sigma$), and 
derived 95\% flux upper limits otherwise. The fluxes (or flux upper limits) and TS values are provided
in Table~\ref{tab:tab1}.  
We also estimated the systematic uncertainties for the spectral data points due to the Galactic diffuse
emission model, by repeating the likelihood analysis in each energy band with the 
normalization of the diffuse component artificially fixed to the $\pm$6\% deviation
from the best-fit value (see e.g., \citealt{abdo+2009,abdo+3-2010}).
The uncertainties estimated in this way are provided in Table~\ref{tab:tab1}, which have been considered together with the statistic ones in Figure~\ref{fig:spectra}.

\section{Discussion and Summary}
\label{sec:disc}

Having analyzed 7 years of \fermi\ LAT Pass 8 data, 
we found excess $\gamma$-ray emission at the \snr\ region. 
As shown in Figure~\ref{fig:tsmap} and \ref{fig:spectra}, both the position
and spectrum (i.e., the high-energy data point at 30 GeV; 
see Table~\ref{tab:tab1}) match those of the TeV emission from the NE shell 
region of \snr\ (see \citealt{ace+10} for details). The detected excess 
emission may well be the GeV counterpart to \snr\ that has been previously 
searched \citep{af12,ace+15}.  
We tested to repeat the analysis in \citet{ace+15}, where
6 years of P7REP data were used, and obtained
the same non-detection result as theirs. Then as we changed to repeat our above analysis using
Pass 8 data in the same 6 years time period, the $>$1 GeV excess emission was detected with
TS$\simeq 20$. Therefore our detection
of the source should be due to the overall sensitivity improvement 
in the Pass 8 data. 

\begin{figure}
\centering
\includegraphics[scale=1.3]{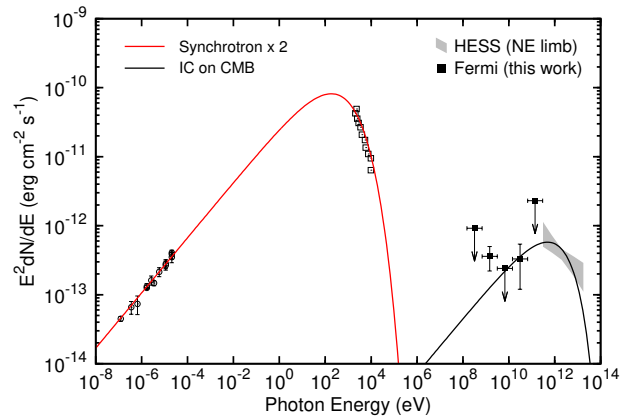}
\caption{Leptonic model fit to the broadband SED of \snr. 
Radio \citep{apg01}, Suzaku X-ray \citep{bam+08}, HESS \citep{ace+10} data 
are included. The solid curve is the model spectra
when $\alpha_e=2.2$.
\label{fig:model}}
\end{figure}

\subsection{Model Fitting}
\label{subsec:mf}

Considering the \fermi\ source as the GeV counterpart and combining
its spectrum with the HESS TeV one, we checked if a leptonic model
that has been proposed (see \citealt{ace+10,af12,ace+15})
could decribe the broadband 
spectrum of \snr. We employed a simple one-zone stationary model,
in which the synchrotron and IC emission originates from the same 
population of electrons with a power-law form plus an exponential cutoff.
Given the radio spectral index of $\sim$0.6 (e.g., \citealt{apg01}), 
the electron spectral index was set to $\alpha_{\rm e}=2.2$. 
To fit the data, for which we have included 
radio \citep{apg01}, X-ray \citep{bam+08},
and GeV and TeV \gr, the required parameters were found to be:
the cutoff energy $E_{\rm cut,e} \approx 17$ TeV,
the total electron energy 
$W_{\rm e}(>1{\rm GeV}) \approx 1.6\times10^{47}$\,erg, 
and the magnetic field strength $B_{\rm SNR}$ $\approx$ 24 $\mu$G. 
These parameter values are compatible with those considered
in the previous leptonic models (see \citealt{af12,ace+15}).
The broadband spectral energy distribution (SED) and the model spectrum
are shown in Figure~\ref{fig:model}, where 
because the radio and X-ray data were emission from the whole remnant, 
the synchrotron 
flux was multiplied by a factor of 2 in our calculation assuming
a symmetry between the NE and SW parts for simplicity. 
Not surprisingly as indicated by the previous studies, the leptonic model generally
can describe the broadband data.
The observed flux at 1.5 GeV is higher than the model flux, but within the 2$\sigma$ uncertainty.

\subsection{Summary}

Analysis of the lastest \fermi\ LAT Pass 8 data for the \snr\ region results
in the detection of a source at a $\sim 4\sigma$ significance
level in 0.15--300\,GeV.
The spectrum of the excess emission can be described by a $\Gamma=1.9$ power 
law (the dashed line in Figure~\ref{fig:spectra}), which can be connected to the HESS spectrum. 
Both the quasar QSO J1504$-$4152 \citep{wl97} and the VHE
source HESS J1504$-$418 \citep{ace+10} are within the 2$\sigma$ error circle of the best-fit position of the source. We searched for variability
in the source, but no significant variations were found. In addition the spectral index of 1.9 does not favor the association to AGN since AGN 
generally have soft power-law spectra with photon indices up to $\sim$3.0 in 
the LAT \gr\ energy range \citep{3fagn15}. 
All of these indicate the source is more likely the GeV counterpart to \snr.

The combined \fermi\ LAT and HESS spectrum can be described by a leptonic model with reasonable parameters, although the data point at 1.5 GeV is slightly higher than our model spectrum. The discrepancy may suggest
that a more complicated model, e.g., multi-emission zones, would be needed
in order to better fit the broadband SED. On the other hand,
we note that the index $\Gamma=1.9\pm0.3$ 
is compatible with those of the other 4 
(RX J1713.7$-$3946, RX J0852.0$-$4622, RCW~86, 
and HESS J1731$-$347; $\Gamma \sim 1.5$) 
TeV shell SNRs summarized by \citet{ace+15}. This similarity may
provide additional evidence for supporting the \gr\ excess emission
as the counterpart to the SNR.
At the source distance of 2.2 kpc, the 0.15--300\,GeV luminosity
is 1$\times 10^{33}$ erg s$^{-1}$, estimated
from the best-fit model. The luminosity value is consistent 
with the general emissional property for the currently \fermi\ detected SNRs:  
young SNRs (with ages no
larger than a few thousands of years) have \gr\ luminosities of 
$\sim 10^{33}$--$10^{34}$\,erg\,s$^{-1}$, while middle-aged, dynamically 
evolved SNRs
are two orders of magnitude brighter due to their interaction with nearby
molecular clouds (see, e.g., \citealt{xin+15} and references therein).

Finally, \citet{mic+14} have reported that 
the SW limb of \snr\ is interacting with
an atomic cloud, and predicted a 3--30 GeV flux of 
$5\times 10^{-13}$ erg\,cm$^{-2}$\,s$^{-1}$ based on their hadronic model
proposed to explain the TeV emission. The flux is reached with the current
\fermi\ LAT data (see Figure~\ref{fig:spectra}). However in our analysis,
we did not see any significant sources in the SW region of \snr\ 
(see Figure~\ref{fig:tsmap}). A simple
estimation for its flux can be used by scaling the NE flux with
the ratio between the HESS SW and NE fluxes (0.67; \citealt{ace+10}),
which gives 1.5--3.4$\times 10^{-13}$\,erg\,cm$^{-2}$\,s$^{-1}$ and 0.8--3.6$\times 10^{-13}$\,erg\,cm$^{-2}$\,s$^{-1}$ at 1.5 and 30 GeV 
(Table~\ref{tab:tab1}), respectively. With
the \fermi\ LAT observation time increasing for the source region, it can
be expected that the SW GeV emission would likely be detected in 
the near future.
The detection will help confirm our result presented here in this work.

\section*{Acknowledgements}

This research made use of the High Performance Computing Resource in the Core
Facility for Advanced Research Computing at Shanghai Astronomical Observatory.
The research was supported by the Shanghai Natural Science 
Foundation for Youth (13ZR1464400), the National Natural Science Foundation
of China for Youth (11403075), the National Natural Science Foundation
of China (11373055), and the Strategic Priority Research Program
``The Emergence of Cosmological Structures" of the Chinese Academy
of Sciences (Grant No. XDB09000000). Z.W. acknowledges the support 
by the CAS/SAFEA International Partnership Program for Creative Research Teams.
Y.C. and X.Z. acknowledge the support of NSFC grant 11233001
and 973 Program grant 2015CB857100.

\begin{table}
\tabletypesize{\footnotesize}
\tablecolumns{4}
\tablewidth{240pt}
\setlength{\tabcolsep}{2pt}
\caption{\fermi\ LAT flux measurements of the source in the region
of \snr}
\label{tab:tab1}
\begin{tabular}{lccc}
\hline
$E$ & Band & $E^2dN(E)/dE$ & TS \\
(GeV) & (GeV) & (10$^{-13}$ erg cm$^{-2}$ s$^{-1}$) \\ \hline
0.32 & 0.15--0.69 & 9.2 & 2 \\
1.47 & 0.69--3.14 & 3.6$\pm$1.3$\pm$0.4 & 10 \\
6.71 & 3.14--14.35 & 2.4 & 1 \\
30.68 & 14.35--65.60 & 3.3$\pm$2.1$\pm$0.0 & 10 \\
140.29 & 65.60--300.00 & 23 & 3 \\
\hline
\end{tabular}
\vskip 1mm
\footnotesize{Note: fluxes with uncertainties are given in energy bins with 
$> 3\sigma$ detection significance, and fluxes without uncertainties are 
the 95$\%$ upper limits. The first and second uncertainties are statistic and systematic ones, respectively.}
\end{table}

%\bibliography{transient}

\end{document}